\documentclass[12pt,a4paper]{article}
\usepackage{latexsym, amsmath}
\renewcommand{\theequation}{\thesection.\arabic{equation}}
\newlength{\extraspace}
\newlength{\extraspaces}
\newcommand{\be}{\begin{equation}
\addtolength{\abovedisplayskip}{\extraspaces}
\addtolength{\belowdisplayskip}{\extraspaces}
\addtolength{\abovedisplayshortskip}{\extraspace}
\addtolength{\belowdisplayshortskip}{\extraspace}}
\newcommand{\ee}{\end{equation}}
 
\newcommand{\ba}{\begin{eqnarray}
\addtolength{\abovedisplayskip}{\extraspaces}
\addtolength{\belowdisplayskip}{\extraspaces}
\addtolength{\abovedisplayshortskip}{\extraspace}
\addtolength{\belowdisplayshortskip}{\extraspace}}
\newcommand{\ea}{\end{eqnarray}}
 
\newcommand{\bas}{\begin{eqnarray*}
\addtolength{\abovedisplayskip}{\extraspaces}
\addtolength{\belowdisplayskip}{\extraspaces}
\addtolength{\abovedisplayshortskip}{\extraspace}
\addtolength{\belowdisplayshortskip}{\extraspace}}
\newcommand{\eas}{\end{eqnarray*}}
 
\newcounter{subequation}[equation]
\makeatletter
\expandafter\let\expandafter\reset@font\csname reset@font\endcsname
\def\subeqnarray{\arraycolsep1pt
    \def\@eqnnum\stepcounter##1{\stepcounter{subequation}
        {\reset@font\rm(\theequation\alph{subequation})}}\eqnarray}

\makeatother
 
\newenvironment{theorem}[1]
{\vspace{3mm}\noindent {\bf #1 :} }{\vspace{2mm}}
\newcommand{\bt}[1]{\begin{theorem}{#1}}
\newcommand{\et}{\end{theorem}}
 
\newcommand{\newsection}[1]{
\vspace{12mm}
\pagebreak[3]
\addtocounter{section}{1}
\setcounter{equation}{0}
\setcounter{subsection}{0}
 
\begin{flushleft}
{\large\bf \thesection. #1}
\end{flushleft}
\nopagebreak
\medskip
\nopagebreak}

\newcommand{\appsection}[1]{
\vspace{12mm}
\pagebreak[3]
\addtocounter{section}{1}
\setcounter{equation}{0}
\setcounter{subsection}{0}
\addcontentsline{toc}{section}
{\protect\numberline{\Alph{section}} {#1}}
 \begin{flushleft}
{\large\bf Appendix:~~#1}
\end{flushleft}
\nopagebreak
\medskip
\nopagebreak}

\newcommand{\la}{\rightarrow}

\newcommand{\is}{\! & \! = \! & \!}

\newcommand{\pa}{{\partial}}
\newcommand{\pp}[1]{ \frac{\partial}{\partial{#1}} }
\newcommand{\dd}[1]{ \frac{\delta}{\delta {#1}} }

\newcommand{\mbar}{{\bar{m}}}

\newcommand{\thetabar}{{\bar{\theta}}}
\newcommand{\Abar}{{\bar{A}}}
\newcommand{\Labar}{{\bar{\Lambda}}}

\newcommand{\cW}{{\cal W}}

\newcommand{\intd}{\int \! d^4 x \;}
\newcommand{\intS}{\int \! d S \;}
\newcommand{\intSbar}{\int \! d \bar S \;}
\newcommand{\intV}{\int \! d V \;}

\newcommand{\equ}[1]{\begin{gather} #1 \end{gather}}
\newcommand{\equa}[1]{\begin{align} #1 \end{align}}

\def\a{\alpha}
\def\be{\beta}

\def\la{\lambda}

\def\Ga{\Gamma}

\def\Gacl{\Ga_{cl}}
\def\La{\Lambda}
\def\dota{{\dot \alpha}}

\def\citere#1{\mbox{Ref.~\cite{#1}}}
\def\citeres#1{\mbox{Refs.~\cite{#1}}}

\def\draftdate{\relax}
\def\mda{\relax}
\def\mua{\relax}
\def\mla{\relax}
\def\draft{
\def\thtystars{******************************}
\def\sixtystars{\thtystars\thtystars}
\typeout{}
\typeout{\sixtystars**}
\typeout{* Draft mode!
         For final version remove \protect\draft\space in source file *}
\typeout{\sixtystars**}
\typeout{}
\def\draftdate{\today}
\def\mua{\marginpar[\boldmath\hfil$\uparrow$]%
                   {\boldmath$\uparrow$\hfil}%
                    \typeout{marginpar: $\uparrow$}\ignorespaces}
\def\mda{\marginpar[\boldmath\hfil$\downarrow$]%
                   {\boldmath$\downarrow$\hfil}%
                    \typeout{marginpar: $\downarrow$}\ignorespaces}
\def\mla{\marginpar[\boldmath\hfil$\rightarrow$]%
                   {\boldmath$\leftarrow $\hfil}%
                    \typeout{marginpar: $\leftrightarrow$}\ignorespaces}
\def\Mua{\marginpar[\boldmath\hfil$\Uparrow$]%
                   {\boldmath$\Uparrow$\hfil}%
                    \typeout{marginpar: $\Uparrow$}\ignorespaces}
\def\Mda{\marginpar[\boldmath\hfil$\Downarrow$]%
                   {\boldmath$\Downarrow$\hfil}%
                    \typeout{marginpar: $\Downarrow$}\ignorespaces}
\def\Mla{\marginpar[\boldmath\hfil$\Rightarrow$]%
                   {\boldmath$\Leftarrow $\hfil}%
                    \typeout{marginpar: $\Leftrightarrow$}\ignorespaces}
\overfullrule 5pt
\oddsidemargin -2mm
\marginparwidth 29mm
}

\begin{document}
%
\begin{titlepage}
%
\renewcommand{\thefootnote}{\fnsymbol{footnote}}
\begin{flushright}
BN-TH-99-09\\
hep-th/9907120
\end{flushright}
\vspace{1cm}
 
\begin{center}
{\Large {\bf  Non-renormalization theorems without supergraphs:}} \\[4mm]
{\Large {\bf The Wess--Zumino Model}}
{\makebox[1cm]{  }       \\[1.5cm]
{\large R. Flume} and {\large E. Kraus  }\\ [3mm]
{\small\sl Physikalisches Institut, Universit\"at Bonn} \\
{\small\sl Nu\ss allee 12, D--53115 Bonn, Germany}}
\\
\vspace{3cm}
{\bf Abstract}
\end{center}
\begin{quote}
The non-renormalization theorems of chiral vertex functions are  derived
on the basis of an algebraic analysis. The property, that the interaction
vertex  is a second supersymmetry variation of a lower dimensional field
monomial,  is
used to relate chiral Green functions to superficially
 convergent Green functions
by extracting the two supersymmetry variations from an internal vertex
and transforming them to derivatives acting on  external legs. 
The analysis is valid in the massive as well as in the massless model and
can be performed irrespective of properties of the superpotential at
vanishing momentum. 
\end{quote}
\vfill
\renewcommand{\thefootnote}{\arabic{footnote}}
\setcounter{footnote}{0}
\end{titlepage}
%
\newsection{Introduction}

Cancellations of ultraviolet divergencies in globally supersymmetric 
field theories, sometimes denoted as mysterious, have been observed at a
very early stage in the development of the field \cite{WeZu74} 
and have been ever since
considered as a hallmark of these theories. Iliopoulos and Zumino
\cite{IlZu74} were
the first to analyze such cancellations, which are not (direct) consequences
of the supersymmetric 
 Ward identities, in the framework of supersymmetric Wess--Zumino models.
Since then several authors \cite{FuLa75, GrRoSi79, GrSi82}
 have contributed to the development
 and streamlining of supergraph formalism in superspace, 
by which the cancellations in various models can be verified straightforwardly.
A careful examination on the other hand of the renormalization of
supersymmetric  field
theories by means of the algebraic renormalization procedure has been
undertaken in \cite{PiSi86}. But  the cancellations of ultraviolet 
singularities, alias the non-renormalization theorems, were  
rather elusive in this approach. Our purpose is to overcome 
this shortcoming by a derivation of the non-renormalization
theorems adaptable to the algebraic renormalization method. As a first step
 into this direction we reconsider here the d=4 Wess--Zumino model. We will show
 that the non-renormalization theorems 
 in this model can be deduced from the fact that  the interaction
Lagrangian of mass dimension 4 may be represented as second supersymmetry 
 variation
of a field monomial with mass dimension 3:
 \begin{equation}
\label{coh}
 F \varphi^ 2 - \frac 12 \varphi \psi^2  
 =  \frac 1 {12} \delta^ {Q}_\alpha \delta^{Q\alpha}
 \varphi^ 3 ,
\qquad \quad
 \bar F \bar\varphi^ 2 - \frac 12 \bar\varphi \bar\psi^2  
 =  \frac 1 {12}\delta^{\bar Q\dota} \delta^ {\bar Q}_\dota 
\bar \varphi^ 3 . 
\end{equation}
Here $\delta^{ Q}_\alpha $ and $\delta^{\bar Q}_\dota $ are the supersymmetry
transformations and $(\varphi, \psi, F)$ is the chiral field multiplet and
$(\bar \varphi, \bar \psi, \bar F)$ is the antichiral field multiplet.

 This sort of cohomological  argument was first used by Zumino \cite{Zu75}
to demonstrate 
that the vacuum graphs in supersymmetric field theories vanish. It was later on
exploited \cite{FlLe84}  for the construction of the perturbative
 (inverse) Nicolai map.

Our argumentation will rely on two ingredients. First of all we extend the
Wess--Zumino model,
 coupling the chiral and anti-chiral interaction vertices to external 
field multiplets. Local couplings have been already considered in the context
of conformal symmetry breaking in the ordinary $\varphi^ 4$ model
\cite{KrSi93}. In the Wess--Zumino model local couplings are necessarily
superfields
and in this way the cohomological structure of the interaction
vertices (\ref{coh})
is included in the action of the extended Wess--Zumino model.
 The original model is recovered in the limit of constant
external field. 
The non-renormalization theorems will appear as simple consequence
of the supersymmetry Ward identities of the extended model.
 We make secondly use
of a modified $R$-symmetry of the extended model involving a non-trivial
$R$-transformation of the external field multiplet. The modified $R$-symmetry 
 will
turn out to be an efficient bookkeeping device for our purposes.

The massless Wess--Zumino model requires a separate consideration. It was noted
in \citeres{West91, JaJoWe91} that in contradiction to  standard lore
 the radiative corrections
to the superpotential of the massless Wess--Zumino model do not vanish.
Deriving the non-renormalization theorems from the cohomological structure
of the interaction vertices,
it is possible to disentangle the ultraviolet and infrared aspects of this
phenomenon.  The supersymmetric specific ultraviolet cancellation based on
eq.~(\ref{coh})  survives the massless limit. The integrated
$\bar \varphi^ 3$-insertion 
of mass dimension 3 on the other hand develops an infrared singularity at 
zero momentum which is canceled by the explicitly extracted powers of
external momenta leaving behind a non-vanishing contribution to the effective
potential.

In section 2 we introduce  local superfield couplings
 in the Wess--Zumino model and
define in this way the Green functions with  covariant vertex insertions.
From the supersymmetry Ward identities the   non-renormalization
theorem for chiral Green functions with one  antichiral vertex insertion 
is derived.
In section 3 we    discuss consequences of the modified $R$-symmetry
and supersymmetry for 1PI Green functions.
In section 4 we relate the
chiral Green functions with antichiral vertex insertions
to the chiral Green functions of the Wess--Zumino model. We profit from
the modified $R$-symmetry, which simplifies the combinatorics and  makes
possible
to generalize the result to arbitrary normalization conditions of
the wave-function renormalization.
 In section 5 we treat the massless model and discuss the
limit to vanishing momenta. In an appendix we give the results
of the algebraic analysis of non-renormalization theorems 
 in expressions of component fields.
Throughout the paper we will use the notations
of \citere{PiSi86}.


\newsection{The Wess--Zumino model with local couplings}

The Wess--Zumino model contains the chiral superfield $A(x, \theta,
\thetabar )$ 
and its complex conjugate, the antichiral  superfield
 $\bar A (x,\theta, \bar \theta)$. In the chiral and antichiral
representation they are expanded in the component fields according to
\begin{equation}
 A (x, \theta) = \varphi(x) + \theta \chi (x) + \theta^ 2 F(x) ,
\qquad
\bar A (x, \bar \theta) = \bar \varphi(x) + \bar\theta \bar \chi (x) + 
\bar \theta^ 2 \bar F(x).
\end{equation}
The classical action\footnote{Throughout the paper we take chiral fields 
in the chiral and antichiral fields  in antichiral representation.}
 \begin{equation}
\label{GaclWZ}
\Ga _{cl} = \frac 1 {16}\int \! d V A
e^ {2i \theta \sigma \thetabar \pa }
\bar A + \frac 1 4 \int \! d S 
\Bigl( \frac m 2 A^2 + \frac \lambda {12}  A ^3 \Bigr) +
\frac 1 4 \int \! d \bar S 
\Bigl( \frac m 2 \Abar^2 + \frac \lambda {12}  \Abar ^3 \Bigr)
\end{equation}
  is invariant under supersymmetry transformations:
\equa{
\label{deltasusy}
\delta _\a ^ Q A  &  = 
 \frac {\partial} {\partial \theta^ \alpha} A   \qquad &
\delta _ \a ^ { Q}\bar  A & =   2 i ( \sigma ^ \mu \bar \theta)_{\alpha}
\partial _\mu \bar A  \\ 
\delta _ \dota ^ {\bar Q} A &  =  - 2 i (\theta \sigma ^ \mu)_{\dota}
\partial _\mu A
\qquad &
\delta _\dota ^  {\bar Q} \bar A & = - \frac {\partial} {\partial\bar \theta^
  \dota} \bar A \nonumber
}
The Wess--Zumino model is renormalizable and   the  Ward identities of
 supersymmetry, 
\begin{equation}
{\cal W}^ {Q}_\alpha \Ga (A,\bar A) = 0\quad \mbox{and}\quad
{\cal W}^ {\bar Q}_{\dot \alpha} \Ga (A, \bar A)
= 0 ,
\end{equation}
define the Green function to all orders of perturbation theory. 
$\Ga(A, \bar A)$ denotes the generating functional of one particle irreducible 
(1PI) Green functions.
 Free parameters are fixed by the following normalization conditions:
\begin{subeqnarray}
\label{normcondWZa}
\Ga_{F \bar F} (p^2) \Big|_{p^2 = \kappa^ 2} = 1 \\
\label{normcondWZb}
\Ga_{F\varphi} (p^2) \Big|_{p^2 = 0} = m \\
\Ga_{F\varphi \varphi} (p_1, p_2, p_3)\Big|_{p_i = 0}  =  - \frac \lambda 2 
\label{normcondWZc}
\end{subeqnarray}
The wave-function renormalization is fixed at an arbitrary normalization point
$\kappa^ 2$, whereas  the mass and 
the coupling  are normalized at $p_i= 0$.
When  the model is constructed with local couplings 
 in section 3 and section 4, 
it is seen, that the normalization point $p_i= 0$ for chiral vertices 
is  indeed 
distinguished from other normalization  momenta.
In this paper
 we refer
 to the BPHZ scheme \cite{BPHZ} in superspace \cite{ClPiSi77}
for removing the divergencies. It has the
advantage to preserve supersymmetry in the subtraction procedure.
 The subtraction degree and the
possible degree of divergence are given by simple power counting theorems in
superspace. In particular the maximal degree of divergence 
for a supergraph with $N_S =  N_A + N_{\bar
  A }$ external legs is determined by
\equa{\label{pcWZ}
d_{ \Omega} \leq & \;
 3 - N_S \quad \mbox{if the supergraph has only chiral or antichiral  legs,} \\
d_{\Omega} \leq & \; 2 - N_S \quad 
\mbox{if  chiral and antichiral legs are present.} \nonumber
}
By  power counting all diagrams belonging to vertices of the classical action 
could be   logarithmically divergent. The non-renormalization theorems of
chiral vertex functions, i.e.~the actual finiteness of all
chiral and antichiral vertex functions,
 are not an immediate consequence of supersymmetry, but
are proven  by an explicit evaluation of diagrams
in superspace \cite{FuLa75, GrRoSi79, GrSi82}.

In the present paper we  derive the non-renormalization theorems
of chiral vertices  from the algebraic property, that the
interaction vertices of the Wess--Zumino model are  supersymmetry variations
of lower dimensional field monomials:
\begin{eqnarray}
\label{chirvar}
 F \varphi^ 2 - \frac 12 \varphi \psi^2  \is   \frac 14
\delta_\a^ {Q} (
 \varphi^2 \psi^\a) =  \frac 1 {12} \delta^ {Q}_\alpha \delta^{Q\alpha}
 \varphi^ 3, \\
\bar  F \bar \varphi^ 2 - \frac 12 \bar \varphi \bar \psi^2  \is   \frac 14
\delta^ {\bar Q\dota} (
 \varphi^2 \bar \psi_\dota) =  \frac 1 {12} \delta^ {\bar Q\dota} \delta^{\bar
   Q}_{\dota }
\bar  \varphi^ 3. \nonumber
\end{eqnarray}
 In  terms of  superfields  these properties are summarized in the 
supersymmetry transformations:
\equa{
\label{chirvarsup}
 \pp {\theta_\a}\pp {\theta^ \a} A ^ 3 & = 
 \delta^{Q \a}  \pp {\theta^ \a} A^ 3 =
\delta^ { Q\a}\delta ^{ Q}_ \a  A^ 3,  \\
 \pp {\thetabar^ \dota}\pp {\thetabar_ \dota} \bar A ^ 3 & = 
-  \delta^{\bar Q}_ \dota  \pp {\thetabar_ \dota} \bar A^ 3 =
\delta ^{\bar Q} _{\dota} \delta^ {\bar Q\dota}\bar  A^ 3.  \nonumber
}
When one wants to consider the effect of this property 
on  Green functions,
 supersymmetric covariant vertices $A^ 3(x, \theta)$
and $ \bar A^ 3 (x, \bar \theta) $ instead of 
the invariant 
 vertices 
$\intS A^ 3(x ,\theta)$ and $
\intSbar \bar A ^ 3 (x,
\bar \theta)$ have to be included into the construction of the Wess--Zumino
model.

The construction of covariant insertions in quantum
field theory is a well understood subject and is most easily performed by
coupling the covariant field polynomials to external fields 
  in such a way,
that the complete action is invariant.  Accordingly we  introduce
external chiral and antichiral superfields with dimension zero,
\begin{equation}
\Lambda (x, \theta ) = \lambda (x) +
\theta \chi(x) + \theta^ 2 f(x),  \qquad
\bar \Lambda (x, \theta ) = \bar \lambda (x) +
\bar \theta \bar \chi(x) + \bar \theta^ 2 \bar f(x), 
\end{equation}
which transform under supersymmetry transformations as the fields $A$ and
$\bar A $ (\ref{deltasusy}). Since the field monomials
$A^ 3 $ and $\bar A^
3$ are transformed as the fields, supersymmetry is maintained, when we
couple $\Lambda $ and  $\bar \Lambda$ to the 
 covariant insertions: 
\begin{equation}
A^ 3 (x, \theta) \longrightarrow \int \! dS \Lambda (x, \theta) A ^3 (x,
\theta),  \qquad 
\bar A^ 3 (x, \bar \theta) \longrightarrow \int \! d\bar S \bar \Lambda (x, \thetabar) 
\bar A ^3 (x, \thetabar ) \nonumber.
\end{equation}
Instead of adding a further interaction term 
to the action of the Wess--Zumino model
we interpret the fields $\Lambda (x, \theta)$ and $\Lambda (x, \bar \theta )$
as local couplings in  superspace:
\begin{equation} 
\label{Gacl}
\Ga _{cl} = \frac 1 {16}\intV A e^ {2i\theta \sigma \thetabar \pa}
\bar A + \frac 1 4 \intS \bigl( 
 \frac m 2  A^2 +\frac 1 {12} \Lambda A ^3 \bigr)
+ \frac  14 \intSbar \bigl( \frac m 2  \bar A ^2 + 
 \frac 1 {12} \bar \Lambda \bar A ^3 \bigr)  .
\end{equation}
The classical action of 
Wess--Zumino model (\ref{GaclWZ}) is recovered by taking the limit
$$\Lambda (x, \theta) \to \lambda \qquad \mbox{and} \qquad
\bar \Lambda (x, \bar \theta)
\to \lambda .$$

Contributions to Green functions  with local couplings are defined 
in higher orders
as usual by the Gell-Mann--Low formula, Wick's theorem and a subtraction
scheme for removing the divergencies.  The BPHZ scheme in superspace
 can be applied to the Wess--Zumino model with local coupling without
modifications. In this scheme the  Ward identities 
of supersymmetry are fulfilled to all
orders by construction:
\begin{equation}
\label{WIsusy1}
{\cal W}^ Q_\a \Ga (A, \bar A, \Lambda, \bar \Lambda )= 0\; ,
 \qquad {\cal W}
^ {\bar Q}_{\dota}
\Ga (A, \bar A, \Lambda, \bar \Lambda ) = 0 
\end{equation}
with
\equa{
\label{WIsusy}
{\cW}_\a^ Q  = &  \intS  \Bigl( \pp {\theta^ \a} A \dd {A} +
 \pp {\theta^ \a} \La \dd
{\La} \Bigr)  +
     \intSbar
 \Bigl( 2i (\sigma ^ \mu \thetabar)_\a \bigl(\pa_\mu \bar A  \dd {\bar A} +
 \pa_\mu \Labar  \dd {\Labar} \bigr)\Bigr) 
 \\
{\cW}_\dota^{\bar Q}  = &  \intSbar  \Bigl(- \pp {\thetabar^ \dota} \bar A
 \dd {\bar A} -
 \pp {\thetabar^ \dota} {\Labar} \dd
{\Labar} \Bigr) 
  + \intSbar \Bigl( - 2i (\theta \sigma ^ \mu )_\dota 
\bigl(\pa_\mu A  \dd A +
 \pa_\mu \La  \dd \La \bigr)\Bigr) \nonumber }
$\Ga = \Ga(A, \bar A, \Lambda, \bar \Lambda )$ is the generating functional
of 1PI Green functions in the Wess Zumino model with local field couplings.
Its lowest order 
$\Ga^ {(0 )}(A, \Abar, \La, \Labar)$ is defined by the classical action
(\ref{Gacl}).
In loop order $L \geq 1$ it is an expansion in external fields $\Lambda,
 \Labar $ and fields $A,  \Abar$: 
\begin{eqnarray}
\label{GaLa}
& & \Ga^ {(L)}(A,\Abar, \La, \Labar) \\
& = & \sum_n \sum_{\bar n} 
\sum_{m = n} ^ {n +2 (L-1)} 
\prod_{k=1}^{n} \int \! d S_k A(z_k)
\prod_{l= 1}^{ \bar n} \int \! d S_l \Abar(\bar z_l) 
\prod_{i=n+1}^{m+n} \int \! d S_i \La(z_i)
\prod_{j=\bar n + 1}^{ \bar n + \bar m}  \int \! d S_j \La(\bar z_j )
\nonumber \\
& & \phantom{\sum \sum \sum \sum}
 \frac 1 {n! m! \bar n! \bar m!} \Ga^ {(L)}_{n,\bar n , m, \bar m} 
(z_1, \ldots z_{m+n}, 
\bar z_1, 
 \ldots \bar z_{\bar m + \bar n})  \nonumber
 \end{eqnarray}
with $z_i \equiv (x_i, \theta_i ) $ and $\bar z_j \equiv (\bar  x_j ,
\thetabar_j) $.
Since
 the perturbative expansion is an expansion in the local couplings,
 the total number of chiral and antichiral vertices is
  determined by the number of
$A$- and $\Abar$-legs and increases by 2 from order to order:
the number $\bar m$ of
antichiral $\Labar$'s is therefore 
not an independent quantity in eq.~(\ref{GaLa}):
\begin{equation}
\label{mbar}
 \mbar \equiv \mbar(n,\bar n, m, L) 
 = n + \bar n + 2(L-1)- m.
\end{equation}
 The generating functional 
of 1PI Green functions of the ordinary Wess--Zumino model is obtained
 in the limit to constant coupling:
\begin{equation}
\lim_{\La, \Labar \to \lambda} \Ga(A, \Abar, \Lambda, \Labar) = 
\Ga (A, \Abar).
\end{equation}
For defining the Green functions of the extended Wess--Zumino model
unambiguously 
 the possible counterterms depending on local couplings  have to be
fixed by suitable normalization conditions and symmetries. We postpone this
discussion to the next section and deduce here immediately the most important
results of the construction. For this purpose 
the normalization conditions are implicitly defined by
the BPHZ scheme at zero momentum.

Differentiating the generating functional with local couplings
(\ref{GaLa}) with respect to $\Lambda $ or
$\Labar$
and taking the limit to constant coupling it is possible study
vertex functions 
of the Wess--Zumino model with one
insertion of a  chiral or antichiral supersymmetric covariant vertex:
\equa{
\label{ins}
 \lim_{\La, \Labar \to \la } \dd {\Lambda (x,\theta)}
 \Ga (A, \Abar, \La, \Labar) 
= & \, \frac 1 {48} [ A^ 3 (x, \theta)]_3 \cdot \Ga (A, \bar A) \\
 \lim_{\La, \Labar \to \la } \dd {\Labar (\bar x,\thetabar)}
 \Ga (A, \Abar, \La, \Labar) 
= & \, \frac 1 {48} [ \bar A^ 3 (\bar x, \thetabar)]_3 \cdot \Ga (A, \bar A)
 \nonumber
 }
In the BPHZ scheme the degree of divergence of a Green function with insertion
$[Q]_d$
   is determined
by the (minimal)  subtraction degree $d$ of the vertex  and the dimensions of
external legs. In superspace  $\theta$'s are counted with dimension $-\frac 12$
and can be extracted from the insertion by raising its dimensions by $+ 
\frac 12$. (See \citere{PiSi86} for a detailed definition of normal products in
 superspace.)
Since the model with local couplings satisfies the supersymmetry
 Ward identities  (\ref{WIsusy1}) one derives consequences for the
Green functions with covariant vertex insertions by differentiating
the Ward identity with respect to
$\Labar$:%
\footnote{From now on we restrict the discussion to 
insertions of antichiral vertices.
 The analog results for chiral insertions
are derived by complex conjugation.}
\begin{eqnarray}
  {\cW}^ {\bar Q}_\dota \dd {\Labar ( x, \thetabar)} \Ga
& = & - \pp {\thetabar^ \dota} \dd {\Labar( x,\thetabar) } \Ga .
\end{eqnarray}
In  the limit to constant $\lambda $ one derives the Ward identity of
supersymmetric covariant vertex insertions:
\begin{eqnarray}
\label{susyins}
  {\cW}^ {\bar Q}_\dota \Bigl([  \bar A^ 3 (x,\thetabar)]_3\cdot \Ga \Bigr)
& = & - [ \pp {\thetabar^ \dota} \bar A^ 3 (x,\thetabar)]_
\frac 52 \cdot \Ga 
\end{eqnarray}
(The Ward operator of supersymmetry on the left-hand-side 
only involves the fields $A$ and $\Abar$.)
Eq.~(\ref{susyins}) is the key relation for
the foundation of the non-renormalization theorems, since it
generalizes  the cohomological property, that invariant vertices 
are supersymmetry variations, eq.~(\ref{chirvarsup}),
to insertions into nonlocal diagrams. Differentiating eq.~(\ref{susyins})
once more with respect to $\thetabar$, one is able to derive an identity, which
relates the highest 4-dimensional 
component of the supermultiplet $\bar A ^ 3$ with
its lowest 3-dimensional component 
by applying twice the Ward operator of supersymmetry
transformations:
\begin{equation}
\label{susyins2}
{\cW_\dota^ {\bar Q}} {\cW^ {\bar Q \dota }} \Bigl([  \bar A^ 3
(x,\thetabar)]_3\cdot \Ga \Bigr) =  
[ \pp {\thetabar^  \dota}
\pp {\thetabar_ \dota} \bar A^ 3 (x,\thetabar)]_4 \cdot \Ga 
\end{equation}
Evaluating this identity for $n$ external chiral  $A(x_i, \theta_i) $-legs
one derives
  consequences  for chiral vertex functions
with an antichiral vertex insertion:
\equa{
&             -  \Bigl([\frac 14 (\bar F \bar 
\varphi^ 2 -\frac 12 \bar \varphi \bar \psi^ 2)]_4  
                        \cdot \Ga\Bigr)_n (x;
 x_1, \theta_1, \ldots x_n,\theta_n ) \\
= & \sum_{k=1}^ n \sum_{l=1}^ n  \Bigl(\theta_k \theta_l \eta^ {\mu\nu} 
                       -i \theta_k \sigma^ {\mu \nu} \theta_l \Bigr)
               \pp {x_k^  \mu}  \pp {x_l^  \nu} 
             \Bigl(\bigl[ \frac 1{12} 
\bar  \varphi^ 3 \bigr]_3 \cdot \Ga\Bigr)_n(x; x_1, \theta_1, \ldots x_n,
\theta_n) \nonumber
}
Integrating over $x$ and using translational invariance
chiral vertex functions with an integrated internal (supersymmetric)
antichiral vertex
are related to 
 vertex functions with a 3-dimensional integrated
$ \bar \varphi^ 3$-insertion and  two explicit
 $x$-derivatives:
\equa{
\label{nonren1}
& - \Bigl(\bigr[\intd \frac 14( \bar F \bar 
\varphi^ 2 -\frac 12 \bar \varphi \bar \psi^ 2
)\bigr]_4 
                     \cdot \Ga\Bigr)_{n}( x_1, \theta_{1n},  \ldots x_n )   \\
= & \sum_{k=1}^ n \sum_{l=1}^ n  \Bigl(\theta_{kn} \theta_{ln}
\eta^ {\mu\nu}  -i \theta_{kn}
 \sigma^ {\mu \nu} \theta_{ln}
                \Bigr) \pp {x_k^  \mu}  \pp {x_l^  \nu}
             \Bigl(\bigl[\intd  \frac 1 {12} \bar  \varphi^ 3 \bigr]_3
 \cdot \Ga\Bigr)_{n}( x_1, \theta_{1n}, \ldots x_n)  \nonumber
}
It is seen from this equation 
that chiral vertex 
functions with one antichiral vertex vanish 
at zero momentum and
by power counting  the
 degree of divergence  is improved by 2 from the   left-hand-side to
the right-hand-side: One gets
one degree of improvement from the dimension of the 3-insertion and 
one degree from  two $\theta$ variables counting each with dimension $\frac
12$.   For the specific normalization conditions 
of the BPHZ scheme
at zero momentum eq.~(\ref{nonren1}) already implies the non-renormalization
theorems of chiral vertex functions in the Wess--Zumino model.
Since any chiral vertex function necessarily includes antichiral vertices,
the 1PI Green functions of the Wess--Zumino model
can be composed  from Green functions with
integrated antichiral vertex insertions, if one takes into account the
correct combinatorial factors.  Due to a new
 $R$-symmetry of the extended Wess--Zumino model it
 turns out that the combinatorics indeed becomes very simple, since in
each loop order the number of antichiral vertices in the contributing
diagrams is  fixed.
In section 4 we complete the above argumentation and generalize the
result in such a way, that also   non-trivial normalization conditions of 
the wave-function renormalization are taken into account.

For completeness we want to mention that on the basis of (\ref{susyins})
and (\ref{susyins2})
one can  study
insertions of antichiral vertices in Green functions with antichiral legs
or with chiral and antichiral legs. In both cases the power counting degree
is not improved by  the relations of the covariant insertions
 since in the supersymmetry transformations no $x$-derivatives are extracted,
but the dimension of external legs is lowered by two $\theta$-differentiations.

\pagebreak


\newsection{$R'$-symmetry, supersymmetry and  local couplings}

The classical action of the Wess--Zumino model with local coupling 
(\ref{Gacl}) is also invariant under a $R$-symmetry,
\begin{equation}
{\cW}^ {R'} \Gacl =0, 
\end{equation} whose generator is
formally given by
\begin{eqnarray}
\label{Rprime}
{\cW}^ {R'} \is \intS i (-A + \theta \pp \theta A) \dd A +
              \intS i (\La + \theta \pp \theta \La) \dd \La \nonumber \\
\!&\!+  \! & \!
\intSbar i (\Abar + \thetabar \pp \thetabar \Abar) \dd \Abar +
              \intSbar
 i (-\Labar + \thetabar \pp \thetabar \Labar) \dd \Labar .
\end{eqnarray}
(We call it $R'$-symmetry, in order to distinguish it from
usual conformal $R$-symmetry of the massless Wess--Zumino model.) 
The generator (\ref{Rprime}) attaches an $R'$-weight $-1$ to the chiral
fields $A$ and a $R'$-weight $+1$ to the local coupling $\Lambda$ with  the
corresponding sign reversal for the respective conjugate fields. 
$\theta$'s and $\thetabar $'s are 
counted with a weight $+1$ and $-1$ respectively. $R'$-symmetry is naively
maintained in the procedure of renormalization.
The BPHZ scheme
adopted further on does respect  supersymmetry and
so we proceed to derive consequences of those symmetries for the quantum action
$\Ga$ (\ref{GaLa}):
\begin{equation}
\Ga = \Gacl + \sum_{L=1}^ \infty \Ga^ {(L)}
\end{equation}
The implementation of supersymmetry restricts the $\theta $- and $\thetabar$-%
dependence of the 1PI Green functions \cite{FuLa75}:
\equa{\label{1PIsusy}
 & \Ga^ {(L)}_{n,\bar n,m,\bar m} 
(x_1, \theta _1, \ldots x_{n+m}, \theta_{n+m},
{\bar x}_1, \thetabar _1, \ldots \bar x_{\bar n + \bar m}, \thetabar _{
\bar n + \bar m}) \\
 &  = 
\exp \Bigl( 2i \sum_{i=1}^{m+n}\theta_i \sigma^ \mu \bar \theta_{\bar n}
 \pa_\mu^ i
    -2i \sum_{j=1}^{\bar n+\bar m } 
\theta_n \sigma^ \mu \thetabar_{j \bar n}\pa_\mu^
    j\Bigr)  
 F_{n,\bar n,m,\bar m}(x_l, \theta_{ln}, \bar x_{\bar l},
 \theta_{\bar l  \bar n})
 \nonumber
}
with
 $ \theta_{in} \equiv
\theta_i - \theta_{n}$ 
 and 
$\thetabar_{j\bar n} \equiv \thetabar_j -
\thetabar_{\bar n}.$  (If only chiral or only antichiral legs and vertices
  are present,
 the exponent vanishes and the respective Green functions depend only on
the differences $\theta_{in}$ or  $\thetabar_{j\bar n}$.)

We want to find out the restrictions due to $R'$-neutrality of the various
contributions. Let us consider a superfield Green function with a fixed
number, say $x$ and $\bar x$,
of  external $\theta$'s and $\thetabar$'s (figuring in the expansions of the
fields $A$ and $\Lambda$ on one hand and of fields $\bar A$ and $\bar \Lambda$
on the other hand).
The coefficient functions $F_{n, \bar n, m , \bar m} $ can
be expanded into a finite series of terms ordered with respect
to the powers $\omega$ and $\bar \omega$ of the difference 
variables $\theta _{im} $ and $\theta_{j \mbar}
$:
\begin{equation}
\label{omegapower}
F_{n,\bar n,m,\mbar}=
\sum_{\omega= 0}^ {2(m+n-1)} \sum_{\bar \omega = 0} ^ {2(\bar n + \bar m -1)}
P^{\omega, \bar \omega}(\theta_{ln}, \thetabar_{\bar l \bar n })
 f_{\omega, \bar
  \omega } (x_l, \bar x_{\bar l} )
\end{equation}  
An additional source of $\theta $'s and $\thetabar$'s is supplied by the
exponential factor in (\ref{1PIsusy}). Let $k$ be the (equal) number of 
$\theta$'s and $\thetabar$'s to a specific term coming from the exponential.
The overall number of $\theta$'s and $\thetabar$'s (internal and external) 
is determined by the number of external legs and vertices. This yields
\begin{equation}
\label{xdet} 
\omega + x + k = 2(m+n) \quad\mbox{and}\quad \bar \omega + \bar x + k = 2(\bar m+\bar n).
\end{equation}
$R'$-neutrality, that is the vanishing of the $R'$ weight of the specific 
contribution,
\begin{equation}
\label{Rneutr}
({\cW}^ {R'} \Ga)_{n,\bar n, m,\mbar} = 0 ,
\end{equation}
is guaranteed through the relation
\begin{equation}
\label{omega}
n + \bar n + 3 m - 3 \mbar - \omega + \bar \omega = 0.
\end{equation}
In addition we have the graphological constraint (\ref{mbar})
\begin{equation}
\label{mbar2}
m + \mbar = n + \bar n + 2(L-1).
\end{equation}
Eliminating in eqs.~(\ref{xdet}), (\ref{omega}) and (\ref{mbar2})
$\omega$ and $\bar \omega$ and solving for $m $ and $\bar m$ one finds
\equa{\label{result}
m & = n + (L -1 ) + \frac 12 (\bar x -x )  \\
\bar m & = \bar n + (L -1 ) - \frac 12 (\bar x -x ).  \nonumber
}
It means in words that the number of chiral and antichiral vertices
 is uniquely 
determined by the setting of external fields and the loop order. It is also
easy to verify that the conditions 
(\ref{xdet}), (\ref{omega}) and (\ref{mbar2})
 cannot be matched non-trivially, if
only chiral legs and chiral vertices are present ($\bar n =  \bar m = 0$ and
$\bar \omega = 0$ ) or if only antichiral legs and antichiral vertices
are present ($ n =   m = 0$ and
$ \omega = 0$ ).

\pagebreak


\newsection{The non-renormalization of chiral vertex functions}

For the consistent renormalization of the Wess-Zumino model with local
 couplings one has to fix all conceivable counterterms
by suitable  normalization conditions.
In the BPHZ scheme possible counterterms are included in  $\Ga_{eff}$:
\begin{eqnarray}
\label{Gaeff}
\Ga_{eff} &= &\Ga_o + \Ga_{int} + \Ga_{ct} 
\end{eqnarray}
with
\begin{eqnarray}
\label{Gaeff2}
\Ga_o & = &\intV  A e ^{2i \theta \sigma \thetabar\pa }
\Abar+ \frac m 8 (\intS A^ 2 + \intSbar \Abar ^ 2)  \\ 
\Ga_{int} & = & \frac 1 {48}\Bigl(\intS \La A^ 3 + \intSbar \Labar\Abar ^
3\Bigr) \nonumber
\end{eqnarray}
and $\Ga_{ct}$ denoting the most general collection of local counterterms
(see below). The dimension of $\Ga_{eff}$ is restricted by renormalizability
to be not larger than 4.
Since the BPHZ scheme   in superspace is a supersymmetric invariant scheme,
the possible counterterms appearing in $\Ga_{eff}$ are supersymmetric.
 As we have shown in the last section
$R'$-invariance excludes Green functions with  exclusively consist of
chiral vertices and chiral legs. Therefore it also forbids to add the
corresponding trivial counterterms to the interaction vertex and to
the mass term:
\begin{subeqnarray}
\label{rcount3}
{\cW }^ {R'} \intS \La ^ n A^ 3 & =& i (n -1) \intS \La^ n A^ 3 \neq  0 ,\;
\mbox{if} \quad n> 1, \\
\label{rcount2}
{\cW }^ {R'} \intS \La ^ n A^ 2 & =& i n \intS \La^ n A^ 2 \neq  0 ,\;
\mbox{if} \quad n> 0.
\end{subeqnarray}
Local field polynomials to the classical chiral vertex functions 
 which include chiral 
$\Lambda$ and antichiral  $\Labar$ have dimension 5 and 6 and do not
 contribute to  $\Ga_{eff}$. Of course, this already expresses, that the
corresponding diagrams are non-local.
 One remains with the counterterms to the kinetic
term, whose dependence on $\Lambda$ and $\Labar$ is again restricted
by $R'$-neutrality (cf.~(\ref{result}) with $\bar x - x = 0$):
\equa{
\label{Gact}
\Ga_{ct} =  \sum_{L=1}^ \infty
 \intV     z^{(L)}  \La ^{L} A e^ {2i \theta \sigma \thetabar
\pa}( \Labar ^ L\Abar     )
}
 We want to mention
that 
there are also some further  counterterms, which 
vanish in the limit to constant coupling. They are not relevant to the present
analysis and we omit them from  $\Ga_{eff}$.
The parameters $z^ {(L)}$ may be fixed in loop order $L$ by the normalization
condition of the Wess-Zumino model (cf.~(\ref{normcondWZa})):
\begin{equation}
\label{normgen}
 \lim_{\lambda(x), \bar \lambda(x) \to \lambda}
\Ga_{F \bar F} (p^2 = \kappa^2 ) = 1
\end{equation}

In order to complete the analysis of non-renormalization theorems of chiral
vertex functions we consider 1PI Green functions with $n$ external chiral
$A$-legs and discuss the limit to constant coupling. Eventually the
chiral Green functions of the Wess-Zumino model are expressed in terms
of  Green functions of the extended Wess-Zumino model, which are convergent
by power counting. The manifest supersymmetric
expressions for Green functions (\ref{1PIsusy}) 
with
 $n$ chiral $A$-legs and $m$ chiral and 
$\bar m$ antichiral vertices  is given by
\begin{eqnarray}
\label{1PIchir}
 & & \Ga^ {(L)}_{n,0,m,\bar m} (x_1, \theta _1 \ldots x_{n+m}, \theta_{n+m},
{\bar x}_1, \thetabar _1 \ldots \bar x_{\bar m}, \thetabar _{\bar m}) \\
 & & = 
e^ {2i \sum_{i=1}^{m+n}\theta_i \sigma^ \mu \bar \theta_{\bar m} \pa_\mu^ i
    -2i \sum_{j=1}^{\bar m -1} \theta_n \sigma^ \mu \bar\theta_{j \mbar}\pa_\mu^
    j}  
 F_{n,0,m,\bar m}(x_1, \theta_{1m+n}, \ldots, x_{n+m}, 
\bar x_1, \thetabar_{1 \mbar}, \ldots, \bar x_{\bar m}) . \nonumber
\end{eqnarray}
The power counting degree of divergence can be determined
by the
power counting formula in superspace \cite{ClPiSi77, PiSi86}
$$d_\Ga = 4- N_S - \sum_s N_s d_s + 
\frac 12 (\omega + \bar \omega).$$
$N_S$ is here the number of all chiral and antichiral legs,
\begin{equation}
N_S =  N_A + N_\La + N_\Labar = n+ m + \mbar = 2n +2( L -1),
\end{equation}
and  $d_s$ denotes the dimension
of the different fields,
\begin{equation}
\sum_s N_s d_s = N_A d_A +N_\La d_\La +N_{\Labar} d_{\Labar} = n.
\end{equation}
$\omega $ and $\bar  \omega$ are the degrees of $\theta$'s and $\thetabar$'s
in the power series expansion of the functions $F_{n,0,m,\mbar}$
(\ref{omegapower}):
\equa{\label{powerc}
d_{\Ga_{n,0,m \mbar}} =&  6 -3n-2L + \frac 12 (\omega + \bar \omega)   
                       \leq 2-n
}
The limit to constant coupling is performed by integrating the $m$ chiral
and $\mbar $ antichiral vertices in superspace:
\equa{\label{limit}
&\Ga^ {(L)}_{n,0,m,\bar m} (x_1, \theta _1, \ldots x_{n+m}, \theta_{n+m},
{\bar x}_1, \thetabar _1, \ldots \bar x_{\bar m}, \thetabar _{\bar m}) 
\stackrel{\Lambda, \Labar \to \lambda}\longrightarrow \\
& \;
\frac 1 {m! \mbar !}
\int \! d S_{n+1} \ldots \int \! d S_{n+m}
 \int \! d \bar S_1 \ldots \int \! d \bar S_{\mbar}  
   \Ga^ {(L)}_{n,0,m,\mbar}(x_1, \theta_{1n+m}, 
\ldots x_{n+m}, \bar x_1,\thetabar_{1\mbar}, \ldots \bar x_{\mbar}) 
\nonumber \\
&= \frac 1 {m! \mbar !}
\int \! d S_{n+1} \ldots \int \! d S_{n+m}
 \int \! d \bar S_1 \ldots \int \! d \bar S_{\mbar}
e^  {2i\sum_{i=1}^ {n-1} \theta_{in} \sigma \thetabar_\mbar
  \pa^ i }
F_{n,0, m,\mbar}(x_1, \theta_{1n+m}, \ldots \bar x_{\bar m})
 \nonumber
}
Of course in this limiting procedure the power counting degree (\ref{powerc})
of
the  vertex functions  is not changed.

For saturation of the $\thetabar$-integrations
in (\ref{limit}) the relevant superfield
Green functions have to depend on $2\mbar$ $\thetabar$'s. 
 Since the functions $F_{n,0,m,\mbar} $ (\ref{1PIchir}) only depend on the
differences $\thetabar_{j\bar m}$, they contribute at most with
$2(\mbar -1) $ $\thetabar$'s. 
The expansion of the exponential function in (\ref{1PIsusy}) breaks off at
second order, which yields the two further  $\thetabar$'s. 
Therefore we find, that all non-vanishing contributions have the following
structure in the number of $\thetabar$'s (cf.~(\ref{xdet})):
\begin{equation}
\bar \omega =  2(\bar m -1 ), \qquad \bar x = 0, \qquad k= 2.
\end{equation}
Let us now consider the number of $\theta$'s. The Green functions with
$\omega= 2 (2n + m -1)$ corresponding to $x= 0$ vanish in
the limit to constant coupling. 
Before integration
these Green functions are related to Green functions with $n$ $\varphi$-legs
and a supersymmetric vertex inserted:
\begin{equation}
\Bigl([  \bar F \bar \varphi ^ 2 - \frac 12 \bar \psi ^ 2 \bar \varphi ]_4
 \cdot \Ga\Bigr)_{\varphi \ldots  \varphi} (x; x_1, \ldots x_n),
\end{equation}
and are not present in the Wess-Zumino model due to supersymmetry.
From $R'$-invariance (\ref{result})
we read off
that they correspond to diagrams with $n+L-1$ chiral vertices and
$L-1$ antichiral vertices. Their power counting degree 
 is determined from eq.~(\ref{powerc}) to be $2-n$.
 
The next less divergent type of superfield Green functions  are the ones
with $\omega = 2(n+m-2)$ corresponding to $x+k = 4$ (\ref{xdet}).
In loop order $L$ they contain according to $R'$-neutrality (\ref{result})
$L$ antichiral  vertices $\Labar$ and $n+L-2$ chiral vertices $\Lambda$.
Since in the limit to constant coupling only the second order in the
expansion of the exponential contributes, the resulting integrated
vertex functions have $2(n-1)$ powers of $\theta$'s and are the chiral
vertex functions $\Ga_n$ of the Wess-Zumino model:
\begin{equation}
\Ga^ {(L)}_{n,0,n+L-2,L} 
\stackrel{\Lambda, \Labar \to \lambda}\longrightarrow \Ga^ {(L)}_n
\end{equation}
Their degree of divergence is determined from the power counting formula
in presence of local
field couplings, eq.~(\ref{powerc}). With
\begin{equation}
\frac 12 (\omega + \bar \omega) = n + m+ \mbar -3 = 2n + 2(L-1) -3
\end{equation}
one has
\begin{equation}
\label{pcchiraln}
d_{\Ga_{n,0,n+L-2,L}} = d_{\Ga_n} = 1-n.
\end{equation}
Compared to power counting in the ordinary Wess-Zumino model (\ref{pcWZ})
the degree of
divergence is improved by 2  and all chiral Green functions 
of the Wess-Zumino model are  convergent emerging from convergent
diagrams of the extended Wess-Zumino model with local couplings. Evaluating
(\ref{limit}) they are seen to vanish at zero momentum, since 
the second order of the exponential function yields two
$x$-differentiations:
\equa{\label{Gan}
\Ga_n^ {(L)}(x_1, \theta_{1n}, \ldots 
x_n ) & =   \lambda^ {n + 2(L-1)}
(\prod_{i= 1}^ {n-1} \theta^ 2_{i n}) f^ {(L)}_n(x_1,\ldots x_n) \\
& =   \lambda^ {n + 2(L-1)}
\sum_{i,k =1 } ^ {n-1}\theta_{in} \theta_{kn} \pa^ \mu_i \pa _\mu^ k
\tilde F _{n,0,n+L-2, L}(x_1, \theta_{1n}, \ldots x_n)
\nonumber
}
with
\equa{
\label{tildeF}
\tilde F _{n,0,m,\mbar}(x_1, \theta_{1n}, \ldots x_n)
= & \;
\frac 1 {m! \mbar !}
\int \! d S_{n+1} \ldots \int \! d S_{n+m}
 \int \! d \bar S_1 \ldots \int \! d \bar S_{\mbar }  \\
 & \; \qquad
(-\thetabar^ 2_{\mbar})  F_{n,0,m,\mbar}(x_1, \theta_{1n+m}, \ldots x_{n+m}, 
\bar x_1,\thetabar_{1\mbar}, \ldots \bar x_{\mbar}) \nonumber
}
By differentiation with respect to the  antichiral coupling
the Green functions $\Ga_{n,0,m,\mbar}$ can be related to Green functions
with covariant $\bar A^ 3$-insertions (cf.~(\ref{ins})). In particular
$\tilde F_{n,0,n+L-2,L} $ (\ref{Gan}) may be expressed in terms of Green
  functions having an integrated $\bar \varphi^ 3$-insertion on an internal
  vertex and otherwise interaction vertices of the Wess-Zumino model.
Similarly $R'$-symmetry makes it  possible, to express chiral Green functions
 of the
Wess-Zumino model directly in terms of Green functions with an antichiral
integrated vertex insertion. In summary, the results of the
algebraic analysis of non-renormalization theorems is expressed
in the following simple
formulae in momentum space:
\begin{eqnarray}
\label{nonrenfin}
&   & \Ga^ {(L)}_{n} (p_1, \theta_{1n}, \ldots p_n)  \\
& &  =  \frac \lambda {L } \Bigl(\bigl[\intd \bar Q_4 (x) \bigr]_4 \cdot
\Ga\Bigr)^{(L)}_n (p_1, \theta_{1n},
\ldots p_n) \nonumber \\
& &  =  \frac \lambda {L }  \Bigl(\bigl[- \frac 1 {4 }  \intd ( \delta^{\bar Q}
_\dota \delta^{\bar Q \dota} \bar  
Q_3 (x) )\bigr]_4 \cdot \Ga\Bigr)^{(L)}_n (p_1, \theta_{1n},
\ldots p_n) \nonumber \\
& & = \frac { \lambda} L
\sum_{i,k=    1}^ {n-1}  \theta_{in}\cdot \theta_{kn} \;p^ \mu_k p_{\mu i}
         \Bigl( \bigl[   
\intd \bar Q_3(x) \bigr]_3 \cdot \Ga \Bigr)^{(L)}_n (p_1, \theta_{1n},
\ldots p_n)\nonumber 
\end{eqnarray}
The insertions are defined by the action principle,
which reads in components ($\bar \Lambda = \bar \lambda (x) + \thetabar
\bar \chi(x) + \thetabar^ 2
\bar f(x) $):
\begin{eqnarray}
\lim_{\Lambda, \Labar \to \lambda } \dd {\bar \lambda (x) } \Ga (A,
\Abar, \Lambda ,
\Labar )
& = &  [\bar Q_4 (x)]_4 \cdot \Ga (A, \bar A)\label{defbarQ4}\\
& = & \bigl[- \frac 1 4 
 (\bar F \bar \varphi^ 2 - \frac 12 \bar \psi^ 2 \bar
\varphi) + {\cal O}(\hbar) \bigr]_4 \cdot \Ga (A, \Abar) \nonumber \\ 
\lim_{\Lambda, \Labar \to \lambda } \dd {\bar f (x) } \Ga (A, \bar A, \La,
\Labar )  
& = &  [\bar Q_3 (x)]_3 \cdot \Ga (A, \Abar)  \label{defbarQ} \\
& = &
\bigl[- \frac 1 {12}   \bar \varphi^ 3 + {\cal O}(\hbar) \bigr]_3
 \cdot \Ga (A, \Abar) \nonumber 
\end{eqnarray}
In the BPHZ scheme the insertions are determined by  $\Ga_{eff}$.
Using  the  general normalization conditions  for the wave-function
 normalization  (\ref{normgen}) the insertions are given by
\begin{eqnarray}
\bar Q_4 (x) \is \lim _{\La, \Labar \to \lambda }  \dd {\bar\lambda (x)}
 \Ga_{eff}\\
\is
- \frac 1 4 (\bar F \bar \varphi^ 2 - \frac 12 \bar \psi^ 2 \bar  \varphi) +
\sum_{L=1 } z^ {(L)}L \lambda^{2L-1} (\bar F F - \frac i 2 \pa^ \mu
 \psi \sigma 
 \bar \psi - \bar \varphi \Box  \varphi) \nonumber \\
\bar Q_3 (x) \is
\lim _{\La, \Labar \to \lambda }  \dd {\bar f (x)} \Ga_{eff}  \\
\is
-\frac 1 {12} \bar \varphi^ 3 +  \sum_{L =1}^ \infty L 
z^ {(L)} \lambda ^
{2L-1} F \bar  \varphi    \nonumber
\end{eqnarray}
In passing we notice that the counterterms to the kinetic term
have the same algebraic properties as the interaction vertex (\ref{chirvar}).
 They can
be expressed as second variations of supersymmetry transformations or
as their complex conjugate:
\begin{subeqnarray}
\label{
kinvar1}
\bar F F + \frac i 2 \psi \sigma \pa \bar \psi  - \varphi \Box \bar \varphi
& = &  \frac 1 4 \delta_\a ^ Q \delta ^ {Q\a} \bar F \varphi \\
\label{kinvar2}
\bar F F - \frac i 2 \pa \psi \sigma \bar \psi  - \bar \varphi \Box \varphi
& = & 
 \frac 1 4 \delta^{\bar Q\dota}  \delta ^ {\bar Q}_{\dota} F\ \bar \varphi
\end{subeqnarray}
In the algebraic analysis of the non-renormalization theorems counterterms
to the kinetic term appear
in the same way as the interaction vertices: Insertions of the supersymmetric
invariant   are transformed into the lowest component of the
respective supermultiplet by extracting at the same time two derivatives on
the external legs.
This implies that the full insertion  related to the derivative of the
antichiral
coupling  can be written as a second supersymmetry variation of a
lower dimensional field monomial.


\newsection{Non-renormalization theorems in the massless  model}

The algebraic analysis of non-renormalization theorems in the massless model
is carried out in the same way as in the massive one.
The renormalization of the massless Wess--Zumino model with local field couplings
can be performed by  the massless version of the BPHZ scheme,
the BPHZL scheme \cite{BPHZL} in superspace \cite{ClPiSi77}. 
In addition to the ultraviolet degree of power counting
one  assigns to every field also an infrared degree of 
power counting
\begin{equation}
\dim^ {IR}A = \dim^ {IR} \Abar = 1 \quad \mbox{and} \quad
\dim^ {IR}\La = \dim^ {IR} \Labar = 0
\end{equation}
The $\Ga_{eff}$ will only include counterterms with infrared dimension
greater or equal  4 (cf.~(\ref{Gaeff}),(\ref{Gaeff2}) and (\ref{Gact})):
\begin{eqnarray}
\Ga_{eff} &= &\Ga_o + \Ga_{int} + \Ga_{ct} 
\end{eqnarray}
with
\begin{eqnarray}
\Ga_o & = &\intV  A e ^{2i \theta \sigma \thetabar\pa }
\Abar+ \frac {M(s-1)} 8 (\intS A^ 2 + \intSbar \Abar ^ 2) \\ 
\Ga_{int} & = & \frac 1 {48}\Bigl(\intS \La A^ 3 + \intSbar \Labar\Abar ^
3\Bigr)\nonumber\\
\Ga_{ct}& = & \sum_{L=1}^ \infty
 \intV    z^{(L)}  \La ^{L}A e ^ {2i\theta \sigma \thetabar \pa}
(\Labar^ L \Abar ) \nonumber
\end{eqnarray}
$M(s-1)$ is the auxiliary mass term of the BPHZL scheme with the subtraction
parameter $s$. The massless limit is achieved by taking $s=1$ after all
subtractions have been performed (see e.g.~\cite{PiSi86} for a detailed
discussion of the BPHZL scheme in superspace).
The massless model satisfies
the  supersymmetry Ward identities (\ref{WIsusy1}) and its
Green functions are $R'$-symmetric
(cf.~(\ref{Rprime}) and (\ref{result})):
\begin{equation}
{\cW}^ {Q}_\a \Ga = 0,  \qquad {\cW}_\dota ^ {\bar Q}\Ga  = 0 
 \quad \mbox{and} \quad {\cW}^ {R'}\Ga  = 0 .
\end{equation}
 The normalization condition of the wave-function normalization
 has to be taken at a non-zero normalization point
$\kappa^ 2$ (\ref{normgen}).
It is well-known that the massless Wess--Zumino model is invariant under
 conformal $R$-symmetry. 
For this reason  the  only non-vanishing chiral 1PI Green functions 
are those with  three external chiral $A$-legs.
The construction of chiral $3$-point functions in expressions of
Green functions of the Wess--Zumino model with local couplings 
proceeds as in the massive model and chiral 3-point functions are related
to   superficially convergent Green functions with covariant vertex insertions.
Taking into account the infrared power counting degree of the
insertions, eq.~(\ref{nonrenfin}) can be taken without modifications to
the massless model:
\begin{eqnarray}
\label{nonrenmassless}
&   & \Ga^ {(L)}_{3} (p_1, \theta_{13}, \ldots p_3)  \\
& &  =  \frac \lambda {L } \Bigl(\bigl[\intd \bar Q_4 (x) \bigr]^ 4_4 \cdot
\Ga\Bigr)_3  (p_1, \theta_{13},
\ldots p_3) \nonumber \\
& &  =  \frac \lambda {L } \Bigl(\bigl[- \frac 1 {4 }  \intd ( \delta^{\bar Q}
_\dota \delta^{\bar Q \dota} \bar  
Q_3 (x) )\bigr]^ 4_4 \cdot \Ga\Bigr)_3 (p_1, \theta_{13},
\ldots p_3) \nonumber \\
& & =  \frac { \lambda} L
 \sum_{i,k=    1}^ {2} \theta_{i3} \theta_{k3} p_k p_i
         \Bigl( \bigl[   
\intd \bar Q_3(x) \bigr]^ 3_3 \cdot \Ga)_3 (p_1, \theta_{13},
\ldots p_3)\nonumber 
\end{eqnarray}
The insertions $\bar Q_4$ and $\bar Q_3$ are defined by differentiation with
respect to the local antichiral coupling (see eqs.~(\ref{defbarQ4})
and (\ref{defbarQ})). In the expressions of eq.~(\ref{nonrenmassless}) there
are no off-shell infrared problems, since
 Green functions with a single integrated 3-3-insertion  exist for
non-exceptional
momenta. Therefore
 the ultraviolet degree of power counting is  improved 
as in the massive model by relating the invariant vertices to the lower
dimensional field monomials, and neither an ultraviolet nor an
infrared overall
subtraction has to be performed (cf.~(\ref{pcchiraln})):
\equ{\label{pcchiralm}
d_\Ga = r_\Ga = -2  \quad \mbox{for chiral 3-point functions}.
}

In contrast to the massive model
the conclusion, that chiral Green functions vanish at zero momentum,
is not true in the massless model, since
 the Green functions with integrated $ \bar Q_3$-insertions
have poles in the external momenta. In the above expressions 
(\ref{nonrenmassless}) these poles are canceled by multiplication
with the external momenta resulting in a non-vanishing and
even in higher than 2-loop order logarithmically divergent contribution to
the effective potential at zero momentum. 
A detailed discussion of the infrared behavior is most conveniently
 done 
in terms of 
component fields. We take  equation (\ref{3point}) with $p_3 = 0$:
\equa{\label{3point0}
\Ga_{F\varphi \varphi}^ {(L)} (p_1,- p_1, 0) =  
 \frac { \lambda }
L& \lim_{p_3 \to 0} \biggl(\; p_3 ^ 2 \Bigl(\bigl[ \intd \bar Q_3(x) 
\bigr]^ 3_3\cdot \Ga\Bigr)^ \a_{\ \a F}
(p_1, p_2,p_3) \\
 & \phantom{\lim_{p_3 \to 0}}+ p_1^ 2 \Bigl(\bigl[ \intd 
\bar Q_3(x)\bigr]^ 3_3\cdot \Ga\Bigr)^ \a_{\ \a F}
(p_1, p_3, p_2) \nonumber\\
&\phantom{\lim_{p_3 \to 0}}+ p_1^ 2 \Bigl(\bigl[ \intd 
\bar Q_3(x)\bigr]^ 3_3\cdot \Ga\Bigr)^ \a_{\ \a F}
(p_2,p_3, p_1)\biggr)\nonumber}
By infrared power counting  we conclude, that the  chiral Green function
of the Wess--Zumino model on the
 left-hand-side exists, since the external $\varphi$-leg, which is associated
with 
$p_3$, has infrared dimension 1. In the lowest contributing loop order $L=2$
there are no divergent subdiagrams and the expression is finite 
according to
 the improved power counting 
formula (\ref{pcchiralm}). For this reason and for dimensional reasons
 the value of the chiral 3-point function at $p_3 = 0$
  is    a pure number in 2-loop order
and exists therefore also at the exceptional momentum $p_1 =
p_3 = 0 $.
In  higher orders self-energy diagrams appear as subdiagrams and
have to be subtracted. Then the chiral $3$-point function 
 depends 
on the ratio  $\frac{p_1^ 2}{ \kappa^ 2}$ and is in general logarithmically
infrared divergent
at $p_1 = 0$:
\equa{
\Ga_{F\varphi \varphi}^ {(2)} (p_1,- p_1, 0) = &
\; C  \\
\Ga_{F\varphi \varphi}^ {(L)} (p_1,- p_1, 0) =  & \; f^ {(L)} ({p_1^2 \over \kappa^
  2}) \quad \mbox{for}
\qquad {L\geq 3}\nonumber
}
The Green functions with
the 3-3-insertions appearing on the right-hand-side of eq.~(\ref{3point0}) 
are in general infrared divergent for $p_3 = 0$, since one produces  in this
 way
a second integrated 3-3- or even 2-2-insertion. 
For dimensional reasons they have the form
\begin{equation}
\label{diman}
\Bigr(\bigl[ \intd \bar Q_3(x)\bigr]_3^ 3\cdot \Ga\Bigr)^ \a_{\ \a F}
(p_1, p_2,p_3) = \sum_{i=1}^ 3\frac 1 {p_i ^ 2} g_i(\frac {p_1^ 2}
{\kappa^ 2},\frac {p_2^ 2}
{\kappa^ 2},\frac {p_3^ 2}
{\kappa^ 2})
\end{equation}
with the functions $g_i$ being at most logarithmically divergent at $p_i = 0$.
If the vanishing momentum is associated
with  the external $F$-leg as it happens in the first line of
 eq.~(\ref{3point0}), 
 one produces in principle
 an integrated  2-2-insertion, which corresponds to a 2-dimensional mass
 insertion into a massless diagram. The corresponding integrals
 do not exist and the Green function in (\ref{diman})
 has a pole  even for $p_1 \neq 0$. In its contribution to the chiral
Green functions of the Wess--Zumino model the pole is canceled
by multiplication with $p_ 3^2$.
In the second and third line of eq.~(\ref{3point0})
 the momenta associated with the $\psi _\a$-legs are put to zero.
From the existence of the left-hand-side at $p_3 = 0$ we can
conclude that the corresponding  Green functions 
with integrated $\bar  Q_3$-insertion exist  for
$p_1 \neq 0$: 
\begin{equation}
\Bigl(\bigl[ \intd \bar Q_3(x)\bigr]_3^ 3\cdot \Ga\Bigr)^ \a_{\ \a F}
(p_1, 0,- p_1 ) =  \frac 1 {p_1^ 2} g(\frac {p_1^ 2}
{\kappa^ 2}).
\end{equation}
At zero momentum
all  poles are canceled by the additional  external momentum factors
leaving behind a non-vanishing finite number in 
2-loop order and  a logarithmically divergent expression in orders
$L\geq 3$.

Since in 
2-loop order the chiral 3-point function is a finite integral,
its value at $p_i = 0$, which has been   calculated in
\citere{JaJoWe91}, is  a characteristic number of the Wess--Zumino model.


\newsection{Conclusions}

The algebraic property of interaction vertices to be second supersymmetry
variations of
lower dimensional field monomials
 has been used to derive the non-renormalization
theorems on the basis of algebraic renormalization.
The technical tool for carrying out the analysis is the extension of the
ordinary coupling of the
Wess-Zumino model to an external field multiplet in superspace.
  Local couplings in
the Wess-Zumino model turned out to be a useful technique for getting insight
into the chiral/antichiral vertex structure of different diagrams:
Based on a modified $R$-symmetry we could prove, that all non-local vertex
functions consist of a definite number of chiral and antichiral
vertices. Green functions with only chiral external $A$-legs are 
 superficially convergent by 
power counting in the extended model. Using the supersymmetry Ward identities
explicit expressions for the chiral vertex functions have been derived by
extracting two supersymmetry variations from an internal antichiral vertex
and transforming them to two derivatives acting on external legs.
By performing this analysis in the massless version the cancellation of
ultraviolet divergencies in chiral vertex functions
 can be proven without referring to properties
of the superpotential inflicted with infrared singularities at
 zero momentum.
 
\vspace{1cm}
{\it Acknowledgments} We would like to thank Klaus Sibold for useful
discussions and comments. One of us (R.F.) acknowledges the support by the TMR
Network Contract FMRX-CT 96-0012 of the European Commision.

\begin{appendix}
\appsection{Non-renormalization theorems in component fields}

In this appendix we  give the expressions for the non-renormalization of
chiral vertex functions,
eq.~(\ref{nonrenfin}), in terms of  component fields.
We start with the 2-point function. 
The contributing diagrams have $L$ chiral and $L$ antichiral vertices 
and the function $\tilde {F}_{2,0, L,L}$ (\ref{tildeF})
does not depend on $\theta$
 due to $R'$-symmetry (cf.~section 3 and in particular eq.~(\ref{result})):
\begin{equation}
\tilde {F}_{2,0, L,L} (x_1, x_2, \theta_{12}) = f (x_1, x_2) = f(x_2,x_1)
\end{equation}
One finds in momentum space:
\begin{equation}
\Ga^ {(L)}_{F\varphi}(p_1, - p_1) =   \frac { \lambda }
L p_1^ 2 \Bigl(\bigl[ \intd 
\bar Q_3 (x)\bigr]_3
\cdot \Ga\Bigr)^ {(L)}_{FF}(p_1, -p_1)
\end{equation}
with $\bar Q_3$ being defined by differentiation with respect to the highest
component $\bar f(x)$ of the antichiral coupling (see eq.~(\ref{defbarQ})). 

For the 3-point function  one has $1+L$ chiral and $L$ antichiral
vertices in every loop order and the function 
$\tilde  {F}_{3,0,1+ L,L} (x_1, x_2, \theta_{12}, \theta_{13}) $ (\ref{tildeF})
is of  degree 2 in $\theta_{i3} $.
After having properly symmetrized we find 
\begin{equation}
\tilde {F}_{3,0, L+1,L} (x_1, x_2, x_3, \theta_{13}, \theta_ {12}) = 
 \theta_{12}^ 2 f(x_1, x_2; x_3)  +
 \theta_{13}^ 2 f(x_1, x_3; x_2) +
 \theta_{23}^ 2 f(x_2, x_3; x_1) 
\end{equation}
with
$\theta_{23} = \theta_{13}-
\theta_{12} $
and $ f(x_1,x_2; x_3) = f(x_2,x_1; x_3) $.
Inserting  $\tilde F_{3,0,L+1,L}$ in this form
 into the generating functional of 1PI Green functions
one derives 
 for the 3-point function of component fields 
the following expression  ($p_1 + p_2 + p_3 = 0$):
\equa{
\label{3point}
\Ga_{F\varphi \varphi}^ {(L)} (p_1, p_2, p_3) =  
 \frac { \lambda }
L \epsilon^ {\alpha \beta}&
  \biggl(\; p_3^ 2 \Bigl(\bigl[ \intd \bar Q_3(x)\bigr]\cdot 
\Ga\Bigr)^ {(L)}_{\beta \a F}
(p_1, p_2, p_3) \\
 & + p_2^ 2 \Bigl(\bigl[ \intd 
\bar Q_3(x)\bigr]\cdot \Ga\Bigr)^ {(L)}_{\beta \a F}
(p_1, p_3, p_2) \nonumber\\
&+ p_1^ 2 \Bigl(\bigl[ \intd 
\bar Q_3(x)\bigr]\cdot \Ga\Bigr)^ {(L)}_{\beta \a F}
(p_2, p_3, p_1)\biggr) \nonumber}

From supersymmetry one has 
\begin{equation}
\Ga_{\a \beta \varphi }(p_1,p_2,p_3) =  \frac 12 \epsilon_{\a \beta}
\Ga_{F\varphi\varphi}(p_1,p_2,p_3)
\end{equation}
and
\equa{
\Bigl(
\bigl[\intd \bar Q_3(x) \bigr]\cdot \Ga\Bigr)_{F  F \varphi} (p_1,p_2,p_3) = & \;
\Bigl(\bigl[\intd \bar Q_3(x) \bigr]\cdot \Ga\Bigr)^ \a_{\ \a F} (p_1, p_3,p_2) \\
+ & \;  
\Bigl(\bigl[\intd \bar Q_3(x) \bigr]\cdot \Ga\Bigr)^ \a_{\ \a F} (p_2, p_3,p_1)
\nonumber }
The Green functions with external fermion legs are defined by:
\begin{equation}
\ldots \dd {\psi^ \beta(x_2)} \dd {\psi^ \alpha(x_1)} \Ga 
\Big|_{{\rm{fields}} = 0}= \Ga_{\a \beta
  \ldots } (x_1, x_2,\ldots)
\end{equation}

For the general $n$-point functions the respective equations are more
complicated, since one finds two supersymmetric functions of degree
$2(n-2)$ in $\theta_{in}$, which contribute
to $\tilde F_{n,0,n+L-2,L}$:
\equa{
\tilde {F}_{n,0,n+ L-2,L} (x_1, \theta_{1n}, \ldots x_n)
& = \Bigl( \prod _{i=2}^ {n}\theta^ 2_{in} f_1 (x_1; x_2, \ldots x_n) +
\mbox{$(n-1)$ perm.}\Bigr)\\ 
 & +   \Bigl(\theta_{1n}\cdot\theta_{2n}\prod_{i= 3}^ {n} \theta^ 2_{in}
  f(x_1, x_2; x_3, \ldots x_n)  +  \; \mbox{$(\frac {n (n-1)}2 -1)$ perm.}
 \Bigr) \nonumber
}
and one gets for the chiral $n$-point functions the result:
\equa{
 &\Ga^ {(L)}_{F\varphi \ldots \varphi}(p_1, p_2, \ldots p_n)    \\ 
 = & \;
 \frac { \lambda} L \biggl(\sum_{k=1}^ {n-1} p_k^ 2 \Bigl( 
\bigl[ \intd \bar Q_3(x)\bigr]_3
\cdot \Ga \Bigr)^ {(L)}_{FF\varphi \ldots \varphi} (p_k, p_n; p_1,
 \ldots \hat p_k,
\ldots p_{n-1}) \nonumber \\
 & \; 
\phantom{\frac { \lambda} L}\! +
2 \epsilon^{\alpha\beta}
\sum^ {n-2}_{k=1} \sum_{i= k+1}^ {n-1} p^ \mu _k p_{\mu i}
\Bigl( \bigl[ \intd \bar Q_3(x)\bigr]_3
\cdot \Ga \Bigr)^ {(L)}_{\beta \a F\varphi \ldots \varphi} (p_k, p_i;p_n,
 p_1, \ldots \hat p_k, \ldots  \hat p_i, 
\ldots p_{n-1}) \biggr)\nonumber}
The notation ``$\hat p_i $'' indicates here  that these momenta are omitted.

\end{appendix}
\pagebreak



\end{document}